\documentclass[11pt,a4paper]{article}

\usepackage[utf8]{inputenc}
\usepackage[T1]{fontenc}
\usepackage{lmodern}
\usepackage{amsmath,amssymb,amsthm}
\usepackage{mathtools}
\usepackage{bm}
\usepackage{natbib}
\usepackage{graphicx}
\usepackage{booktabs}
\usepackage{hyperref}
\usepackage[margin=2.5cm]{geometry}
\usepackage{algorithm}
\usepackage{algorithmic}
\usepackage{caption}
\usepackage{subcaption}
\usepackage{xcolor}
\usepackage{tikz}
\usepackage{authblk}
\usetikzlibrary{arrows.meta,positioning,calc}

\hypersetup{
  colorlinks=true,
  linkcolor=black,
  citecolor=black,
  urlcolor=black,
}
\graphicspath{{figures/}}

\newtheorem{remark}{Remark}

\newcommand{\E}{\mathbb{E}}
\newcommand{\Prob}{\mathbb{P}}

\newcommand{\VaR}{\operatorname{VaR}}
\newcommand{\CVaR}{\operatorname{CVaR}}
\newcommand{\GPD}{\operatorname{GPD}}
\newcommand{\Pois}{\operatorname{Poisson}}
\newcommand{\Nor}{\mathcal{N}}
\newcommand{\iid}{\overset{\text{iid}}{\sim}}

\title{Bayesian Extreme Value Theory with Hawkes-AR-Gumbel Dependence\\
for Extreme CVaR Estimation in Operational Risk}

\author[1,2]{Juan Ballesteros G\'omez}
\author[1,3]{Eduardo C. Garrido-Merch\'an}
\author[1,2]{Pedro Pablo P\'erez-Velasco}

\affil[1]{Universidad Pontificia Comillas, Madrid, Spain}
\affil[2]{Banco Santander, Madrid, Spain}
\affil[3]{Instituto de Investigaci\'on Tecnol\'ogica (IIT), Universidad Pontificia Comillas, Madrid, Spain}

\date{May 14, 2026}

\begin{document}

\maketitle

\begin{center}
\small
\texttt{202533373@alu.comillas.edu} \quad
\texttt{ecgarrido@comillas.edu} \quad
\texttt{ppperez@comillas.edu}
\end{center}

\begin{abstract}
Operational risk capital estimation under Basel~II/III requires quantifying
aggregate losses at extreme confidence levels of~99.9\% and beyond, yet the
standard Loss Distribution Approach (LDA) assumes independence between loss
frequency and severity, an assumption frequently violated during stress
episodes when both the number and the size of losses increase simultaneously.
Furthermore, maximum likelihood estimation of tail parameters ignores
parameter uncertainty, leading to overconfident risk estimates at extreme
quantiles. We propose a Bayesian framework that combines Extreme Value Theory
with a dynamic dependence architecture, the Hawkes-AR-Gumbel model, for
operational risk Conditional Value-at-Risk (CVaR) estimation at confidence
levels up to~99.995\%. The model integrates three mechanisms that capture
empirically documented features of operational losses: an autoregressive
latent stress process that captures persistence of crisis regimes, a Hawkes
self-excitation component for frequency that generates event clustering and
overdispersion, and a Gumbel copula for upper-tail dependence that links
frequency and severity innovations through an asymmetric copula concentrating
dependence in the extreme tail. Inference is performed via Hamiltonian Monte
Carlo using PyMC, yielding full posterior distributions for all parameters,
and CVaR at arbitrary confidence levels is estimated through posterior
predictive Monte Carlo simulation. We compare three models on simulated
operational risk data: the independent model (standard LDA), a shared latent
factor model with symmetric dependence, and the proposed Hawkes-AR-Gumbel
model. The independent model underestimates CVaR at~99.995\% by approximately
40\%, while the shared factor model fails to capture temporal persistence,
event clustering, and upper-tail asymmetry. The Hawkes-AR-Gumbel model
recovers the true dependence structure and correctly estimates CVaR at
extreme levels, with the entire computation completing in under 15~minutes
on a standard laptop.

\medskip
\noindent\emph{Keywords:} Operational risk, Extreme Value Theory,
Generalized Pareto Distribution, Bayesian inference, CVaR, Expected
Shortfall, Hawkes process, Gumbel copula, autoregressive latent stress, MCMC,
frequency-severity dependence.
\end{abstract}

\section{Introduction}\label{sec:intro}

Operational risk, the risk of loss resulting from inadequate or failed
internal processes, people, and systems, or from external events, constitutes
one of the three pillars of bank capital regulation under the Basel~II/III
framework \citep{bcbs2006}. Unlike market and credit risk, operational risk is
characterized by low-frequency, high-severity events whose statistical
properties are notoriously difficult to estimate \citep{mcneil2015}. The
standard regulatory approach, the Loss Distribution Approach (LDA), models the
annual aggregate loss as
\begin{equation}\label{eq:lda}
  S = \sum_{i=1}^{N} X_i,
\end{equation}
where $N$ is the annual number of loss events and $X_1, X_2, \ldots$ are the
individual loss severities \citep{frachot2001}. Under the standard LDA, $N$
and the $X_i$ are assumed to be independent, an assumption that simplifies
computation but reflects neither the bursty frequency dynamics nor the
heavy-tailed severities that operational-risk data exhibit, both of which
strain the inference of capital at the 99.9\% quantile \citep{cope2009}.

This paper addresses three interrelated limitations of the standard approach.
The first is the independence assumption itself. The standard LDA treats $N$
and the $X_i$ as mutually independent, which underestimates the aggregate
loss tail when frequency and severity are positively dependent. Moreover, the
dependence between frequency and severity is typically asymmetric: it
concentrates in the upper tail, where crises synchronize frequency and
severity simultaneously, while calm periods exhibit little correlation.
Symmetric dependence structures such as shared Gaussian factors cannot
capture this asymmetry \citep{embrechts2002}. The second limitation is the
absence of temporal dynamics. Operational loss events exhibit temporal
clustering of the self-exciting type \citep{hawkes1971,bacry2015}: large
events tend to arrive in bursts rather than uniformly over time. Furthermore,
the underlying stress environment is persistent, meaning that adverse
conditions in one period propagate into the next rather than dissipating
immediately. Standard LDA
models, with their constant Poisson rate and time-invariant severity
distribution, are structurally unable to capture these dynamics. The third
limitation concerns parameter uncertainty at extreme quantiles. Maximum
likelihood estimation of the Generalized Pareto Distribution (GPD) tail
parameters provides point estimates that ignore estimation uncertainty. At
extreme quantiles ($\geq 99.9\%$), where the data are sparse by definition,
this uncertainty is a first-order concern: small changes in the shape
parameter $\xi$ translate into large changes in the estimated quantile, and
the asymptotic approximations underlying frequentist confidence intervals
break down precisely in the regime where they are most needed
\citep{coles1996,peters2006}.

To address these three limitations simultaneously, we propose the
\emph{Hawkes-AR-Gumbel model}, a Bayesian framework that integrates three
mechanisms. The first is an AR(1) latent stress process
$Z_t = \phi Z_{t-1} + \varepsilon_t$ that captures the persistence of crisis
regimes: with $\phi \approx 0.7$, a stress shock takes approximately three
years to dissipate, reflecting the empirically documented persistence of
operational risk environments. The second is Hawkes self-excitation for the
frequency process, where past events increase the intensity of future events
through
$\lambda_t = \mu(Z_t) + \sum_{s<t} \eta\, N_s\, e^{-\kappa(t-s)}$. This
generates the temporal clustering and overdispersion that a simple Poisson
model cannot reproduce. The third mechanism is a Gumbel copula linking the
frequency and severity innovation processes through an asymmetric upper-tail
dependence structure with coefficient $\lambda_U = 2 - 2^{1/\theta} > 0$.
Unlike the Gaussian copula, for which $\lambda_U = 0$, the Gumbel copula
ensures that extreme frequency and extreme severity co-occur with positive
probability even asymptotically.

Inference is fully Bayesian via Hamiltonian Monte Carlo implemented in PyMC
\citep{salvatier2016}, yielding posterior distributions for all structural
parameters and latent variables. CVaR is estimated at arbitrary confidence
levels through posterior predictive Monte Carlo simulation, naturally
integrating over parametric, latent, and structural uncertainty. Using a
simulation-based calibration approach, we demonstrate that ignoring
frequency-severity dependence underestimates CVaR at~99.995\% by approximately
40\%. We also show that symmetric dependence through a shared Gaussian factor
captures some dependence at moderate quantiles but fails to improve CVaR
estimation at extreme quantiles, precisely because the Gaussian copula lacks
upper-tail dependence. The Hawkes-AR-Gumbel model recovers the true
data-generating process parameters and correctly estimates CVaR across the
full range of confidence levels considered. The entire computation runs in
under~15 minutes on a standard laptop, making the approach practical for
internal model validation and regulatory stress testing.

The contributions of the paper are threefold. First, we propose a unified
Bayesian dependence architecture that combines copula-based upper-tail
dependence, autoregressive persistence, and self-excitation, three mechanisms
that have been studied individually but, to our knowledge, never jointly
within an operational-risk Bayesian framework. Second, we provide a complete
posterior predictive algorithm for CVaR estimation that integrates over all
sources of uncertainty and remains computationally tractable at the extreme
quantiles relevant for regulatory capital. Third, through a controlled
simulation study with known ground truth, we quantify the systematic
underestimation produced by the standard LDA and isolate the contribution of
each dependence mechanism, in particular demonstrating why symmetric
dependence is insufficient at the extreme quantiles where capital adequacy is
determined.

The remainder of this paper is organized as follows.
Section~\ref{sec:lit} reviews the related literature on extreme value theory
for operational risk, frequency-severity dependence modeling, and Bayesian
tail estimation, situating our contribution within each stream.
Section~\ref{sec:methodology} presents the three models in detail,
progressing from the standard independent LDA through the shared latent
factor model to the full Hawkes-AR-Gumbel specification.
Section~\ref{sec:inference} describes the Bayesian inference procedure,
including the rationale behind each prior distribution and the posterior
predictive algorithm for CVaR estimation. Section~\ref{sec:simulation}
presents the simulation study, including the data-generating process,
parameter recovery results, and the central CVaR comparison across models.
Section~\ref{sec:discussion} discusses the regulatory implications of our
findings, explains why symmetric dependence fails at extreme quantiles, and
identifies the conditions under which frequency-severity dependence is most
consequential. Section~\ref{sec:conclusion} concludes with a summary of
contributions and directions for future research. Two appendices collect the
Marshall-Olkin algorithm for sampling from the Gumbel copula and the closed
form of the Gumbel log-density used in the MCMC likelihood.

\section{Related Literature}\label{sec:lit}

Our work sits at the intersection of three literatures: extreme value theory
for operational risk, frequency-severity dependence modeling, and Bayesian
inference for tail estimation. We review each in turn, highlighting both the
contributions we build upon and the gaps our framework addresses.

The use of Extreme Value Theory for modeling operational risk tails was
pioneered by \citet{mcneil1999} and \citet{embrechts1997}, who advocated the
Peaks-over-Threshold (POT) approach based on the Generalized Pareto
Distribution. The theoretical justification for this approach comes from the
Pickands-Balkema-de Haan theorem \citep{pickands1975,balkema1974}, which
establishes that exceedances above a sufficiently high threshold converge in
distribution to the GPD regardless of the parent distribution. In the
operational risk context, GPD shape parameters $\xi$ typically range from
0.3 to~1.5, indicating very heavy tails with infinite variance or even
infinite mean in some cases \citep{chavez2004,moscadelli2004}. An important
development within this tradition was the work of \citet{chavez2004}, who
introduced smooth extremal models that allow GPD parameters to vary with
covariates. \citet{embrechts2003} provided an early synthesis of how extreme
value methods translate into regulatory capital figures, emphasizing that
quantile estimation at the 99.9\% level is qualitatively different from
estimation at less extreme quantiles. Our latent stress factor can be
understood as a generalization of the smooth-extremal idea, where the
covariate is itself a latent process inferred jointly with the tail
parameters.

The standard LDA independence assumption has been challenged on both
empirical and theoretical grounds. \citet{cope2009} documented the severe
data-sufficiency challenges that arise when estimating operational-risk
capital at the 99.9\% quantile from limited loss data, observing that the
required extrapolation can exceed by two orders of magnitude the largest
observed loss and recommending that supervisors consider less extreme
quantiles. More recently,
\citet{garzon2023} proposed skew-$t$ copulas to model asymmetric upper-tail
dependence among severity components across business lines and event types in
operational risk, providing evidence that the dependence structure is
concentrated in the upper tail. Our
contribution differs from these approaches by combining three dependence
mechanisms, copula-based upper-tail dependence, autoregressive persistence,
and self-excitation, in a unified Bayesian framework. While each mechanism
has been studied individually, their combination is, to our knowledge, novel.

Self-exciting point processes, introduced by \citet{hawkes1971}, have found
extensive application in financial modeling, particularly for high-frequency
data. \citet{bacry2015} provide a comprehensive review of Hawkes processes
in finance, documenting their ability to capture event clustering in order
flow, trade arrivals, and default cascades. \citet{ogata1981} provided the
canonical thinning algorithm that underlies efficient simulation of these
processes. \citet{zhu2013} studied ruin probabilities under non-stationary
arrivals with heavy-tailed claims, providing theoretical support for
combining self-excitation with extreme value distributions. Our discrete-time
annual Hawkes component adapts the continuous-time framework to the typical
granularity of operational loss databases, where events are aggregated at
annual or quarterly frequency.

The use of copulas to model dependence structures separately from marginal
distributions is well-established in quantitative risk management, with the
foundational decomposition due to \citet{sklar1959} and modern treatments in
\citet{joe2014} and \citet{mcneil2015}. The Gumbel copula belongs to the
Archimedean family and is the canonical choice for upper-tail dependence,
with coefficient $\lambda_U = 2 - 2^{1/\theta}$ for parameter $\theta \geq
1$ \citep{nelsen2006}. \citet{embrechts2002} issued an influential warning
about the limitations of Gaussian copulas for capturing tail dependence,
showing that the Gaussian copula's zero upper-tail dependence coefficient can
lead to severe underestimation of joint extreme risks. Despite this warning,
the operational risk literature has been slow to adopt asymmetric
alternatives. Our framework makes this transition explicit by replacing the
implicit Gaussian copula of shared-factor models with the Gumbel copula,
which concentrates dependence precisely where it matters most for capital
estimation.

The Conditional Value-at-Risk we adopt as our principal risk measure was
formalized by \citet{rockafellar2002} and shown to satisfy the coherence
axioms of \citet{artzner1999}, in contrast to Value-at-Risk, which lacks
sub-additivity in general. CVaR is therefore the appropriate target for
operational risk capital allocation across business lines and risk cells.
Finally, \citet{coles1996} demonstrated that Bayesian methods for extreme
value modeling provide more honest uncertainty quantification than
frequentist alternatives, particularly at extreme quantiles where asymptotic
approximations break down. \citet{peters2006} applied Bayesian Monte Carlo
methods specifically to operational risk, showing that posterior predictive
distributions for aggregate losses can be substantially wider than
frequentist confidence intervals, and argued that this additional width
reflects genuine epistemic uncertainty that practitioners should not ignore.
Our approach extends this line of work by embedding the Bayesian GPD within
a richer dependence structure that captures temporal dynamics and upper-tail
asymmetry.

\section{Methodology}\label{sec:methodology}

We consider a panel of operational loss data observed over $T$ years. In each
year~$t$, the bank observes $N_t$ loss events with individual severities
$X_{t,1}, \ldots, X_{t,N_t}$ exceeding a threshold~$u$. The exceedances are
$Y_{t,i} = X_{t,i} - u$. The annual aggregate loss is
$S_t = \sum_{i=1}^{N_t} (u + Y_{t,i})$. We present three models of increasing
complexity, all sharing the GPD severity assumption but differing in their
treatment of frequency-severity dependence and temporal dynamics. This
progression from simple to complex serves both a pedagogical purpose, making
the contribution of each mechanism transparent, and a practical one,
providing baselines against which the full model can be evaluated.

\subsection*{The independent model (standard LDA)}\label{sec:model_ind}

The simplest specification assumes complete independence between frequency
and severity. The number of events in each year follows a Poisson distribution
with a constant rate, $N_t \sim \Pois(e^{\mu_\lambda})$, and each exceedance
follows a GPD with constant parameters,
$Y_{t,i} \sim \GPD(e^{\mu_\sigma}, \xi)$. The log-parametrization of the
rate and scale ensures positivity without the need for constrained
optimization. This model has three free parameters,
$\bm{\psi}_1 = (\mu_\lambda, \mu_\sigma, \xi)$. The GPD density for $y > 0$
with $\xi > 0$ takes the form
\begin{equation}\label{eq:gpd_density}
  f(y \mid \sigma, \xi) = \frac{1}{\sigma}
    \left(1 + \frac{\xi y}{\sigma}\right)^{-(1+1/\xi)},
    \qquad 1 + \frac{\xi y}{\sigma} > 0.
\end{equation}
The use of the GPD for exceedances above a high threshold is justified by the
Pickands-Balkema-de Haan theorem, which establishes that the excess
distribution $F_u(y) = \Prob(X - u \leq y \mid X > u)$ converges to the GPD
as the threshold~$u$ approaches the upper endpoint of the distribution, for
any parent distribution in the maximum domain of attraction of an extreme
value distribution \citep{pickands1975,balkema1974}. This result provides a
distribution-free justification for using the GPD in our setting: regardless
of the unknown true loss distribution, the exceedances above a sufficiently
high threshold are approximately GPD-distributed. The independent model
serves as the regulatory benchmark. It is widely implemented, computationally
simple, and well-understood. Its limitation is equally clear: by treating
frequency and severity as unrelated, it is structurally unable to capture the
phenomenon that stress periods produce both more events and costlier events,
a pattern that amplifies the aggregate loss tail beyond what independent
convolution would predict.

\subsection*{The shared latent factor model}\label{sec:model_dep}

The second model introduces a latent annual stress factor $Z_t$ that creates
dependence between frequency and severity:
\begin{align}
  Z_t &\iid \Nor(0, 1), \label{eq:dep_z}\\
  N_t \mid Z_t &\sim \Pois\!\left(\exp(\mu_\lambda + \alpha Z_t)\right),
    \label{eq:dep_freq}\\
  Y_{t,i} \mid Z_t &\sim \GPD\!\left(\exp(\mu_\sigma + \beta Z_t),\; \xi\right).
    \label{eq:dep_sev}
\end{align}
The mechanism is intuitive. When $\alpha > 0$ and $\beta > 0$, a high
realization of the stress factor ($Z_t \gg 0$) simultaneously increases the
Poisson rate $\lambda_t = \exp(\mu_\lambda + \alpha Z_t)$ and inflates the
GPD scale $\sigma_t = \exp(\mu_\sigma + \beta Z_t)$, producing years with
both more events and heavier tails. Conversely, low-stress years
($Z_t \ll 0$) produce fewer, cheaper losses. This model has five structural
parameters, $\bm{\psi}_2 = (\mu_\lambda, \alpha, \mu_\sigma, \beta, \xi)$,
plus $T$ latent variables $Z_t$. The shared Normal factor represents a
genuine improvement over the independent model, but it carries an important
structural limitation. Because $Z_t$ is normally distributed, the induced
dependence between frequency and severity is symmetric: the copula implicit
in this construction is Gaussian, and the Gaussian copula has zero upper-tail
dependence ($\lambda_U = 0$). In practical terms, this means that the
probability of simultaneous extremes in frequency and severity vanishes
asymptotically. The dependence captured by the shared factor is therefore
strongest at moderate quantiles and weakens precisely at the extreme
quantiles where capital adequacy is determined. This limitation motivates
the third model.

\subsection*{The Hawkes-AR-Gumbel model}\label{sec:model_hag}

Our proposed model extends the shared factor framework with three mechanisms
that address the limitations identified above. We describe each component and
its role in the overall architecture. The first component replaces the
implicit Gaussian copula with an explicit Gumbel copula for the innovation
processes. Let $(W_t^f, W_t^s)$ be bivariate innovations with standard normal
marginals, linked by the Gumbel copula
\begin{equation}\label{eq:gumbel_copula}
  C_\theta(u,v) = \exp\!\left(-\left[(-\ln u)^\theta +
    (-\ln v)^\theta\right]^{1/\theta}\right),
    \qquad \theta \geq 1.
\end{equation}
The Gumbel copula belongs to the Archimedean family with generator
$\varphi(t) = (-\ln t)^\theta$. Its defining property for our purposes is
\emph{upper-tail dependence}: the probability that both innovations are
simultaneously extreme remains positive even asymptotically, with coefficient
\begin{equation}\label{eq:lambda_u}
  \lambda_U = \lim_{q \to 1^-} \Prob(V > q \mid U > q)
    = 2 - 2^{1/\theta}.
\end{equation}
For $\theta = 1$ the copula reduces to independence ($\lambda_U = 0$); as
$\theta$ increases, the upper-tail dependence strengthens. For our
application, $\theta = 2$ yields $\lambda_U \approx 0.59$, so that
conditional on the frequency innovation being in its extreme upper tail
there is a~59\% probability that the severity innovation is also extreme.
This is a qualitative departure from the Gaussian copula, where this
probability converges to zero. The copula density, needed for the likelihood
computation, is
\begin{equation}\label{eq:gumbel_density}
  c_\theta(u,v) = \frac{C_\theta(u,v)}{uv} \cdot
    \frac{(\ell_u \ell_v)^{\theta-1}}{A^{2-1/\theta}} \cdot
    \left(A^{1/\theta} + \theta - 1\right),
\end{equation}
where $\ell_u = -\ln u$, $\ell_v = -\ln v$, and
$A = \ell_u^\theta + \ell_v^\theta$. The full derivation of the log-density
used in the MCMC implementation is provided in Appendix~\ref{app:density}.

The second component introduces \emph{autoregressive persistence} in the
latent stress. Rather than drawing $Z_t$ independently each year, we let the
frequency innovation drive an AR(1) process,
\begin{equation}\label{eq:ar1}
  Z_t = \phi Z_{t-1} + W_t^f, \qquad |\phi| < 1.
\end{equation}
The persistence parameter $\phi$ controls how long stress shocks propagate.
With $\phi = 0.7$, the autocorrelation at lag~$h$ is $\rho(h) = \phi^h$, so a
shock takes approximately $-1/\ln\phi \approx 2.8$ years to halve in
magnitude. The stationary distribution is $Z_t \sim \Nor(0, 1/(1-\phi^2))$,
which has variance approximately~1.96 when $\phi = 0.7$, nearly twice the
unit variance of the independent specification. This increased unconditional
variance means that the AR(1) specification produces more extreme stress
realizations than the independent model, compounding the effect on aggregate
losses. The choice of an AR(1) structure is motivated by the empirical
observation that operational risk stress regimes, such as those following a
major fraud discovery or a systemic information-technology failure, persist
across multiple years rather than appearing and disappearing at random.

The third component adds \emph{Hawkes self-excitation} to the frequency
process. The intensity of the Poisson frequency in year~$t$ combines the
stress-driven baseline with a feedback term from past event counts,
\begin{equation}\label{eq:hawkes}
  \lambda_t = \exp(\mu_\lambda + \alpha Z_t) +
    \sum_{s=1}^{t-1} \eta\, N_s\, e^{-\kappa(t-s)},
\end{equation}
where $\eta > 0$ is the excitation amplitude and $\kappa > 0$ is the decay
rate. The frequency count is then
$N_t \mid Z_t, N_{<t} \sim \Pois(\lambda_t)$. The economic intuition is that
a year with many loss events, for example due to the discovery of a
widespread fraud scheme, increases the probability of further events in
subsequent years as investigations widen, copycat behavior emerges, or
systemic vulnerabilities are exposed. The branching ratio
\begin{equation}\label{eq:branching}
  r = \frac{\eta\, e^{-\kappa}}{1 - e^{-\kappa}}
\end{equation}
determines the long-run amplification factor $1/(1-r)$. If $r = 0.3$, then
the self-excitation increases the expected frequency by approximately~43\%
relative to the stress-only baseline. We impose the subcriticality constraint
$r < 0.95$ to ensure that the process remains stationary and does not
explode. Finally, the severity distribution depends on the copula-linked
severity innovation,
\begin{equation}\label{eq:sev_hag}
  Y_{t,i} \mid W_t^s \sim \GPD\!\left(\exp(\mu_\sigma + \beta_s W_t^s),\;
    \xi\right).
\end{equation}
Because $(W_t^f, W_t^s)$ are linked through the Gumbel copula, the scale
parameter $\sigma_t$ is correlated with the frequency innovation $W_t^f$,
and hence with the stress $Z_t$, through upper-tail dependence rather than
symmetric correlation. This is the key mechanism through which the model
generates heavier aggregate loss tails than the shared factor model: in
stress episodes, both the number of events and the cost per event are
simultaneously extreme, with the Gumbel copula ensuring that this
synchronization is strongest precisely in the tail.

\begin{figure}[htbp]
\centering
\begin{tikzpicture}[
    node distance=1.6cm and 2.2cm,
    param/.style={draw, circle, minimum size=1.1cm, fill=blue!8,
      font=\small},
    latent/.style={draw, circle, minimum size=1.1cm, fill=red!8,
      font=\small},
    obs/.style={draw, circle, minimum size=1.1cm, fill=gray!20, thick,
      font=\small},
    copula/.style={draw, circle, minimum size=1.1cm, fill=orange!15,
      font=\small},
    arrow/.style={-{Stealth[length=2.5mm]}, thick, gray!70!black},
    every label/.style={font=\footnotesize},
  ]
  \node[copula] (gumbel) {$C_\theta$};
  \node[latent, below left=1.4cm and 2cm of gumbel] (wf) {$W_t^f$};
  \node[latent, below right=1.4cm and 2cm of gumbel] (ws) {$W_t^s$};
  \node[latent, below=1.6cm of wf] (z) {$Z_t$};
  \node[param, left=1.5cm of z] (phi) {$\phi$};
  \node[obs, below=1.6cm of z] (n) {$N_t$};
  \node[param, left=1.5cm of n] (hawk) {$\eta,\kappa$};
  \node[obs, below=2.8cm of ws] (x) {$Y_{t,i}$};
  \draw[arrow] (gumbel) -- node[above left, font=\scriptsize] {copula} (wf);
  \draw[arrow] (gumbel) -- node[above right, font=\scriptsize] {copula} (ws);
  \draw[arrow] (wf) -- node[right, font=\scriptsize] {innovation} (z);
  \draw[arrow] (phi) -- node[above, font=\scriptsize] {persist.} (z);
  \draw[arrow] (z) -- node[right, font=\scriptsize] {stress} (n);
  \draw[arrow] (hawk) -- node[above, font=\scriptsize] {self-excit.} (n);
  \draw[arrow] (ws) -- node[right, font=\scriptsize] {scale} (x);
  \node[obs, below right=1.4cm and 0.5cm of n] (s) {$S_t$};
  \draw[arrow] (n) -- (s);
  \draw[arrow] (x) -- (s);
  \draw[arrow, red!60!black, dashed]
    (n.west) to[out=180,in=270] node[left, font=\scriptsize,
    text=red!60!black] {feedback} (hawk.south);
  \draw[arrow, blue!60] (z.north west) to[out=135,in=225,looseness=5]
    node[left, font=\scriptsize, text=blue!60] {AR(1)} (z.south west);
\end{tikzpicture}
\caption{Directed acyclic graph of the Hawkes-AR-Gumbel model. Shaded nodes
  ($N_t$, $Y_{t,i}$, $S_t$) are observed. The Gumbel copula $C_\theta$ links
  the frequency and severity innovations $(W_t^f, W_t^s)$, creating
  upper-tail dependence. The AR(1) structure in $Z_t$ creates temporal
  persistence (blue self-loop), and the Hawkes feedback from past counts
  $N_{<t}$ through $(\eta,\kappa)$ to $N_t$ (red dashed) creates event
  clustering.}
\label{fig:dag}
\end{figure}
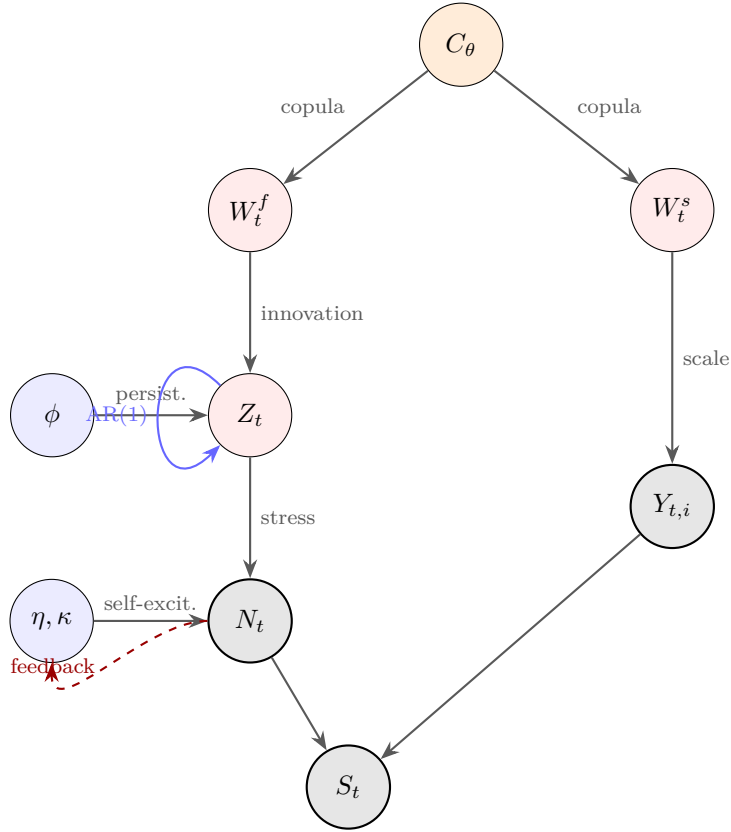

The complete Hawkes-AR-Gumbel model has nine structural parameters,
\[
  \bm{\psi}_3 = (\phi, \mu_\lambda, \alpha, \eta, \kappa,
  \mu_\sigma, \beta_s, \xi, \theta),
\]
plus $2T$ latent innovation variables
$(W_1^f, \ldots, W_T^f, W_1^s, \ldots, W_T^s)$ and $T$ deterministic latent
stress values $Z_1, \ldots, Z_T$. Figure~\ref{fig:dag} displays the
directed acyclic graph summarizing the model's dependence structure.

\begin{remark}\label{rem:nesting}
The three models form a strictly nested hierarchy. The independent model
arises from the shared factor model in the limit $\alpha = \beta = 0$, and
the shared factor model arises from the Hawkes-AR-Gumbel model in the limit
$\phi = 0$, $\eta = 0$, $\theta = 1$. This nesting structure makes each
model a special case of the next, allowing us to attribute observed
differences in CVaR estimation to specific mechanisms rather than to overall
flexibility differences.
\end{remark}

\section{Bayesian Inference and CVaR Estimation}\label{sec:inference}

\subsection*{Prior specification}\label{sec:priors}

We adopt weakly informative priors that regularize the posterior while
allowing the data to dominate. The choice of each prior reflects domain
knowledge about operational risk parameters and is designed to prevent the
sampler from exploring implausible regions of the parameter space without
unduly constraining the posterior. Table~\ref{tab:priors} summarizes all
prior distributions; we discuss the reasoning behind the most consequential
choices below.

\begin{table}[htbp]
\centering
\caption{Prior distributions for the Hawkes-AR-Gumbel model parameters.
  The latent innovations $W_t^f$ and $W_t^s$ are given independent standard
  normal priors and tied through the Gumbel copula via a potential
  correction (see Section~\ref{sec:likelihood}).}
\label{tab:priors}
\begin{tabular}{llll}
\toprule
Component & Parameter & Prior & Justification \\
\midrule
AR(1)     & $\phi$       & $\text{Beta}(5, 2)$               & Mean $\approx 0.71$; stress persists \\
Frequency & $\mu_\lambda$ & $\Nor(3, 1.5)$                    & $e^3 \approx 20$ events/year \\
Frequency & $\alpha$     & $\text{HalfNormal}(1)$            & Positive sensitivity to stress \\
Hawkes    & $\eta$       & $\text{HalfNormal}(0.3)$          & Moderate self-excitation \\
Hawkes    & $\kappa$     & $\text{HalfNormal}(1)$            & Annual decay rate \\
Severity  & $\mu_\sigma$ & $\Nor(14, 2)$                     & $e^{14} \approx 1.2$M \\
Severity  & $\beta_s$    & $\text{HalfNormal}(1)$            & Positive copula effect \\
GPD       & $\xi$        & $\text{TruncNorm}(0.5, 0.5; [0.01, 2])$ & Heavy tail, bounded \\
Copula    & $\theta$     & $1 + \text{HalfNormal}(1.5)$      & $\theta \geq 1$ boundary \\
Latent    & $W_t^f, W_t^s$ & $\Nor(0,1)$                     & Standard normal marginals \\
\bottomrule
\end{tabular}
\end{table}

The persistence parameter $\phi$ receives a $\text{Beta}(5,2)$ prior, which
has mean $5/7 \approx 0.71$ and concentrates most of its mass on values
between~0.4 and~0.95. This reflects a substantive belief, grounded in the
operational risk literature, that stress regimes last multiple years: a bank
that suffers a major fraud event in year~$t$ is likely to face elevated loss
rates in years~$t+1$ and~$t+2$ as investigations unfold, remediation proves
incomplete, and organizational vulnerabilities persist. The Beta family is a
natural choice for a parameter restricted to $(0,1)$, and the $(5,2)$
parameterization tilts the prior toward high persistence without ruling out
lower values if the data demand them. The GPD shape parameter $\xi$ receives
a truncated normal prior centered at~0.5 with standard deviation~0.5,
restricted to the interval $[0.01, 2]$. The centering at~0.5 reflects the
consensus in the operational risk literature that loss tails are heavy
($\xi > 0$), with typical values ranging from~0.3 to~1.5
\citep{moscadelli2004,chavez2004}. The truncation at~0.01 prevents the
pathological case $\xi \leq 0$ (bounded tails), which is empirically
implausible for operational losses, while the upper bound at~2.0 prevents
extreme values that would imply infinite mean losses. The standard deviation
of~0.5 is wide enough to let the likelihood dominate: even with~15 years of
data, the posterior for $\xi$ is substantially narrower than the prior.

The copula parameter $\theta$ is reparametrized as
$\theta = 1 + \theta_{-1}$ with $\theta_{-1} \sim \text{HalfNormal}(1.5)$.
The reparametrization ensures $\theta \geq 1$, since $\theta = 1$ corresponds
to independence. The HalfNormal(1.5) prior places substantial mass near
independence ($\theta \in [1, 2]$) while allowing the data to push toward
stronger dependence ($\theta > 3$) if warranted. This is deliberately
conservative: we prefer to let the data demonstrate the presence of
upper-tail dependence rather than assuming it. The frequency parameters
$\mu_\lambda$ and $\alpha$ have standard weakly informative priors. The
$\Nor(3, 1.5)$ prior on $\mu_\lambda$ is centered at $\ln(20) \approx 3$,
reflecting a baseline expectation of approximately~20 threshold exceedances
per year, a plausible rate for a single operational risk cell. The
HalfNormal(1) prior on $\alpha$ restricts the stress sensitivity to be
non-negative, encoding the economic constraint that stress should not
decrease the frequency of loss events. Similarly, the severity scale
$\mu_\sigma$ receives $\Nor(14, 2)$, centered at $\ln(10^6) \approx 13.8$,
reflecting losses on the order of millions, and the copula effect
$\beta_s \sim \text{HalfNormal}(1)$ is restricted to be positive. The Hawkes
parameters $\eta$ and $\kappa$ receive HalfNormal priors with scales chosen
to keep the branching ratio $r$ moderate; the subcriticality constraint
$r < 0.95$ is enforced via a potential that assigns $-\infty$ log-likelihood
to violations, preventing the sampler from exploring non-stationary regimes.

\subsection*{Likelihood}\label{sec:likelihood}

The complete log-likelihood for the Hawkes-AR-Gumbel model consists of three
additive components,
\begin{align}
  \ell(\bm{\psi}, \bm{W} \mid \bm{N}, \bm{Y}) &=
    \underbrace{\sum_{t=1}^{T} \log p(N_t \mid \lambda_t)}_{\text{Poisson frequency}}
    + \underbrace{\sum_{t=1}^{T} \sum_{i=1}^{N_t}
      \log f_{\text{GPD}}(Y_{t,i} \mid \sigma_t, \xi)}_{\text{GPD severity}}
    \nonumber\\
    &\quad + \underbrace{\sum_{t=1}^{T}
      \log c_\theta\!\left(\Phi(W_t^f), \Phi(W_t^s)\right)}_{\text{Gumbel copula correction}},
    \label{eq:loglik}
\end{align}
where $\Phi(\cdot)$ is the standard normal CDF. The first term is the
standard Poisson log-likelihood with intensity $\lambda_t$ given by
equation~\eqref{eq:hawkes}. The second term is the GPD log-likelihood,
$\log f_{\text{GPD}}(y \mid \sigma, \xi) =
-\log\sigma - (1 + 1/\xi) \log(1 + \xi y / \sigma)$,
with the support constraint $1 + \xi y / \sigma > 0$.

The third term deserves particular attention. The innovation variables
$(W_t^f, W_t^s)$ enter the model with independent standard normal priors. If
these priors were left unmodified, the innovations would be independent, and
the model would reduce to the shared factor specification with the AR(1)
structure. The copula correction term adjusts their joint density from
bivariate independence to the Gumbel copula structure: for each year~$t$, we
compute $u_t = \Phi(W_t^f)$ and $v_t = \Phi(W_t^s)$ and add the log-density
of the Gumbel copula evaluated at $(u_t, v_t)$. This approach, implementing
the copula as a potential correction to independent marginal priors, is
computationally convenient because it avoids the need to sample directly
from the copula distribution within the MCMC, while producing the correct
joint posterior.

\subsection*{MCMC implementation}\label{sec:mcmc}

We implement the model in PyMC~5 \citep{salvatier2016} using the No-U-Turn
Sampler (NUTS), a variant of Hamiltonian Monte Carlo that adaptively selects
the integration path length \citep{hoffman2014,neal2011}. For the
Hawkes-AR-Gumbel model, we use~2 chains with~2000 warmup iterations and~2000
sampling iterations each, with a target acceptance rate of~0.98. The
elevated target acceptance rate, relative to the default of~0.80, is
necessary because the posterior geometry is complex: the Gumbel copula
correction introduces sharp curvature in the log-posterior surface, and the
AR(1) structure creates strong correlations between successive latent
variables. A lower acceptance rate leads to excessive divergent transitions
and poor exploration of the tails. The simpler models use a target
acceptance rate of~0.90, which suffices for their less complex posteriors.
The AR(1) structure in $Z_t$ is implemented as a deterministic scan over the
innovations $W_t^f$, computing $Z_1 = W_1^f$ and
$Z_t = \phi Z_{t-1} + W_t^f$ for $t \geq 2$. This non-centered
parameterization, where the sampler explores the innovation space rather
than the state space, is critical for efficient sampling: centered
parameterizations of latent time series are known to produce strong
posterior correlations between successive states and the persistence
parameter, leading to slow mixing
\citep{papaspiliopoulos2007}. Convergence is assessed using the
split-$\widehat{R}$ diagnostic \citep{vehtari2021} together with bulk and
tail effective sample sizes (ESS). We require $\widehat{R} < 1.01$ and
$\text{ESS}_{\text{bulk}} > 400$ for all structural parameters before
proceeding to CVaR estimation.

\subsection*{CVaR estimation via posterior predictive simulation}\label{sec:cvar_alg}

The CVaR (Conditional Value-at-Risk, also known as Expected Shortfall) at
confidence level~$q$ is defined as
\begin{equation}\label{eq:cvar_def}
  \CVaR_q = \E[S \mid S \geq \VaR_q],
\end{equation}
where $\VaR_q$ is the $q$-quantile of the aggregate loss distribution~$S$
\citep{rockafellar2002,artzner1999}. We estimate CVaR through posterior
predictive simulation, which naturally integrates over all sources of
uncertainty: parametric, latent, and structural. The procedure, detailed in
Algorithm~\ref{alg:cvar}, works by repeatedly drawing a parameter vector
from the posterior, generating a single annual loss scenario under that
parameter vector, and aggregating the results across $M = 10^6$ simulations.
Because each simulation uses a different posterior draw, the resulting
empirical distribution of~$S$ reflects not only the stochastic variability
of losses under fixed parameters, but also the uncertainty about the
parameters themselves.

\begin{algorithm}[htbp]
\caption{Posterior predictive CVaR estimation (Hawkes-AR-Gumbel)}
\label{alg:cvar}
\begin{algorithmic}[1]
\REQUIRE Posterior samples $\{(\phi^{(j)}, \alpha^{(j)}, \mu_\lambda^{(j)},
  \eta^{(j)}, \kappa^{(j)}, \mu_\sigma^{(j)}, \beta_s^{(j)}, \xi^{(j)},
  \theta^{(j)})\}_{j=1}^{J}$, threshold $u$, confidence levels $\{q_k\}$
\ENSURE VaR and CVaR estimates at each level
\FOR{$i = 1$ \TO $M$ (e.g., $M = 10^6$)}
  \STATE Draw parameter index $j \sim \text{Uniform}\{1,\ldots,J\}$
  \STATE Sample $(U, V) \sim C_{\theta^{(j)}}$ via Marshall-Olkin algorithm
  \STATE Set $W^f = \Phi^{-1}(U)$, $W^s = \Phi^{-1}(V)$
  \STATE Sample $Z_{\text{past}} \sim \Nor\!\bigl(0,\, \phi^{(j)2}/(1-\phi^{(j)2})\bigr)$;
    set $Z = Z_{\text{past}} + W^f$
  \STATE Compute $r^{(j)} = \eta^{(j)} e^{-\kappa^{(j)}} / (1 - e^{-\kappa^{(j)}})$
  \STATE Set $\lambda = \exp(\mu_\lambda^{(j)} + \alpha^{(j)} Z) / (1 - r^{(j)})$
  \STATE Sample $N \sim \Pois(\min(\lambda, 500))$
  \STATE Set $\sigma = \exp(\mu_\sigma^{(j)} + \beta_s^{(j)} W^s)$
  \FOR{$k = 1$ \TO $N$}
    \STATE Sample $Y_k \sim \GPD(\sigma, \xi^{(j)})$
  \ENDFOR
  \STATE Set $S^{(i)} = \sum_{k=1}^{N} (u + Y_k)$
\ENDFOR
\FOR{each confidence level $q_k$}
  \STATE $\VaR_{q_k} = \text{quantile}(\{S^{(i)}\}, q_k)$
  \STATE $\CVaR_{q_k} = \text{mean}(\{S^{(i)} : S^{(i)} \geq \VaR_{q_k}\})$
\ENDFOR
\end{algorithmic}
\end{algorithm}

Two aspects of the algorithm merit explanation. First, in line~5, we draw
$Z_{\text{past}}$ from the stationary distribution of the AR(1) process,
$\Nor(0, \phi^2/(1-\phi^2))$, rather than simulating the full time series.
This is valid because we are simulating a representative annual loss from
the stationary distribution, not a specific year in a time series. Second,
in line~7, the stationary amplification factor $1/(1-r)$ replaces the
explicit Hawkes summation. Under stationarity, the expected intensity of the
Hawkes-augmented Poisson is $\lambda_{\text{base}}/(1-r)$, where
$\lambda_{\text{base}} = \exp(\mu_\lambda + \alpha Z)$ is the stress-driven
component. This approximation avoids the need to simulate a multi-year
history for each of the $10^6$ draws, reducing computation from hours to
seconds. The Gumbel copula samples in line~3 are generated using the
Marshall-Olkin algorithm, which proceeds by first drawing a positive stable
random variable $M \sim S(1/\theta, 1)$ via the Chambers-Mallows-Stuck
algorithm, then setting $U = \exp(-(E_1/M)^{1/\theta})$ and
$V = \exp(-(E_2/M)^{1/\theta})$ where $E_1, E_2 \iid \text{Exp}(1)$. The
details are given in Appendix~\ref{app:gumbel}.

\section{Simulation Study}\label{sec:simulation}

\subsection*{Data-generating process}\label{sec:dgp}

We evaluate the three models using simulated data generated from the
Hawkes-AR-Gumbel data-generating process (DGP) with the parameter values
listed in Table~\ref{tab:dgp_params}. This simulation-based calibration
approach \citep{talts2018} allows us to assess both parameter recovery,
whether the posterior concentrates around the true values, and CVaR
estimation accuracy, whether the model produces correct tail risk estimates,
against known ground truth. While simulation cannot demonstrate real-world
applicability, it provides the strongest possible test of internal
consistency and is a prerequisite for any model intended for regulatory use.

\begin{table}[htbp]
\centering
\caption{True parameter values for the data-generating process used in the
  simulation study.}
\label{tab:dgp_params}
\begin{tabular}{llrl}
\toprule
Component & Parameter & Value & Interpretation \\
\midrule
AR(1)     & $\phi$                   & 0.70  & Persistence; shocks halve in $\sim$3 years \\
Frequency & $\mu_\lambda = \ln(20)$  & 3.00  & Base rate $\approx$ 20 events/year \\
Frequency & $\alpha$                 & 0.50  & Moderate stress sensitivity \\
Hawkes    & $\eta$                   & 0.30  & Self-excitation amplitude \\
Hawkes    & $\kappa$                 & 0.50  & Annual decay rate \\
Severity  & $\mu_\sigma = \ln(10^6)$ & 13.82 & Base GPD scale $\approx$ 1M \\
Severity  & $\beta_s$                & 0.40  & Copula effect on scale \\
GPD       & $\xi$                    & 0.70  & Heavy tail (Pareto-type) \\
Copula    & $\theta$                 & 2.00  & $\lambda_U \approx 0.59$ \\
Threshold & $u$                      & $5\times 10^5$ & 500K threshold \\
\bottomrule
\end{tabular}
\end{table}

\begin{figure}[htbp]
\centering
\includegraphics[width=0.95\textwidth,keepaspectratio]{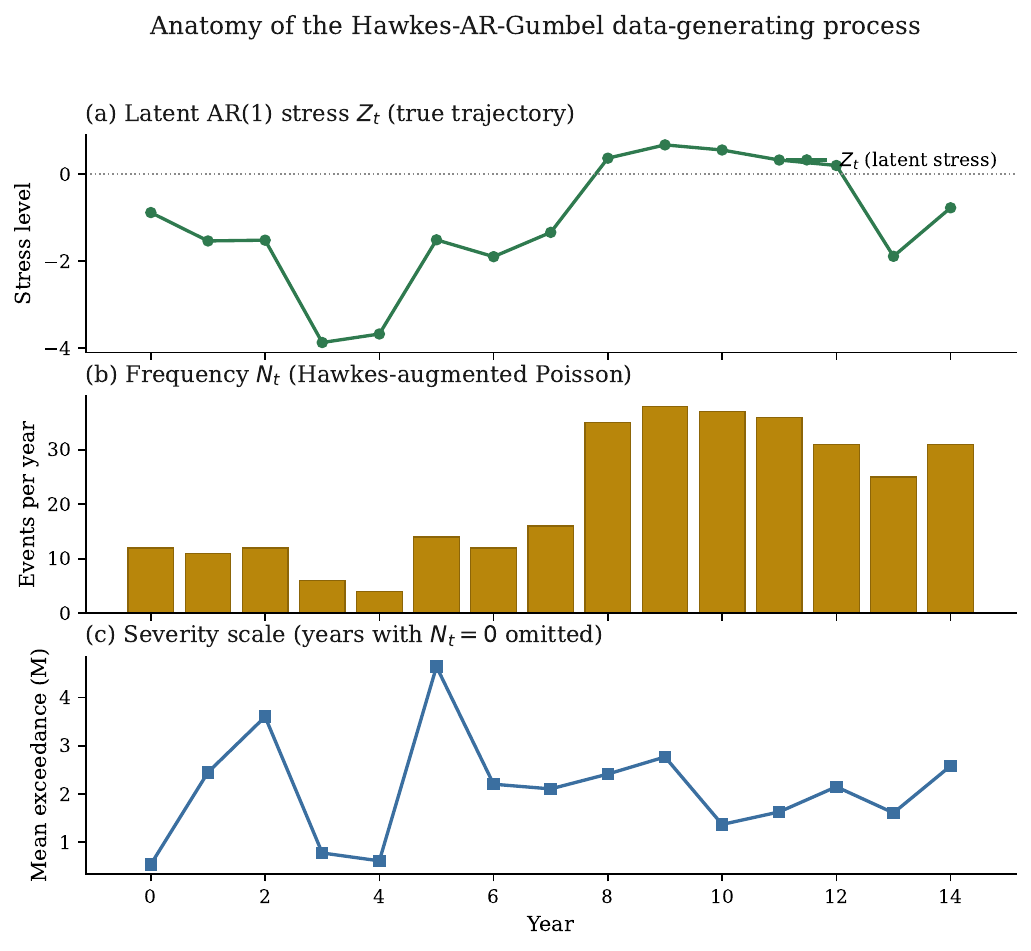}
\caption{Anatomy of the Hawkes-AR-Gumbel data-generating process across the
  15-year simulation window. From top to bottom, the panels display the
  latent stress $Z_t$, the realized event count $N_t$, and the per-year
  mean exceedance. The visible bursts in $N_t$ during high-stress periods,
  together with the simultaneous inflation of the mean exceedance, reflect
  the joint operation of the AR(1) persistence, the Hawkes self-excitation,
  and the Gumbel upper-tail dependence.}
\label{fig:dgp}
\end{figure}

The simulation follows the DGP described in Section~\ref{sec:methodology}.
We first generate copula innovations $(U_t, V_t) \sim C_\theta$ via the
Marshall-Olkin algorithm and transform them to normal scale,
$W_t^f = \Phi^{-1}(U_t)$, $W_t^s = \Phi^{-1}(V_t)$. The AR(1) stress is
then built as $Z_1 = W_1^f$, $Z_t = \phi Z_{t-1} + W_t^f$ for $t \geq 2$.
For each year, the Hawkes-augmented frequency
$\lambda_t = e^{\mu_\lambda + \alpha Z_t} +
\sum_{s<t} \eta\, N_s\, e^{-\kappa(t-s)}$
is computed, event counts $N_t \sim \Pois(\lambda_t)$ are sampled, the
severity scale $\sigma_t = e^{\mu_\sigma + \beta_s W_t^s}$ is determined by
the copula innovation, and exceedances $Y_{t,i} \sim \GPD(\sigma_t, \xi)$
are drawn for each event. The resulting dataset contains $T = 15$ years
with approximately~925 total loss events exceeding the threshold.
Figure~\ref{fig:dgp} illustrates the resulting trajectories of the latent
stress, frequency, and severity processes across the 15-year window.

\subsection*{Parameter recovery}\label{sec:recovery}

Table~\ref{tab:posterior} reports the posterior summary for the key
parameters of the Hawkes-AR-Gumbel model fitted to the simulated data.

\begin{table}[htbp]
\centering
\caption{Posterior summary for the Hawkes-AR-Gumbel model
  (2 chains $\times$ 2000 post-warmup draws). HDI denotes the 94\% highest
  density interval. All $\widehat{R}$ values equal~1.0, indicating
  convergence.}
\label{tab:posterior}
\begin{tabular}{lccccr}
\toprule
Parameter & True & Mean & SD & 94\% HDI & $\widehat{R}$ \\
\midrule
$\phi$ (persistence)         & 0.70  & 0.68  & 0.15 & $[0.38,\; 0.92]$  & 1.00 \\
$\alpha$ (stress $\to$ freq) & 0.50  & 0.39  & 0.14 & $[0.19,\; 0.64]$  & 1.00 \\
$\eta$ (Hawkes excitation)   & 0.30  & 0.23  & 0.17 & $[0.00,\; 0.53]$  & 1.00 \\
$\kappa$ (Hawkes decay)      & 0.50  & 0.60  & 0.41 & $[0.04,\; 1.34]$  & 1.00 \\
$\mu_\sigma$ (log GPD scale) & 13.82 & 13.84 & 0.08 & $[13.69,\; 13.98]$ & 1.00 \\
$\beta_s$ (copula $\to$ sev) & 0.40  & 0.40  & 0.11 & $[0.24,\; 0.62]$  & 1.00 \\
$\xi$ (GPD shape)            & 0.70  & 0.73  & 0.06 & $[0.63,\; 0.84]$  & 1.00 \\
$\theta$ (Gumbel)            & 2.00  & 2.33  & 0.78 & $[1.09,\; 3.76]$  & 1.00 \\
\bottomrule
\end{tabular}
\end{table}

\begin{figure}[htbp]
\centering
\includegraphics[width=0.95\textwidth,keepaspectratio]{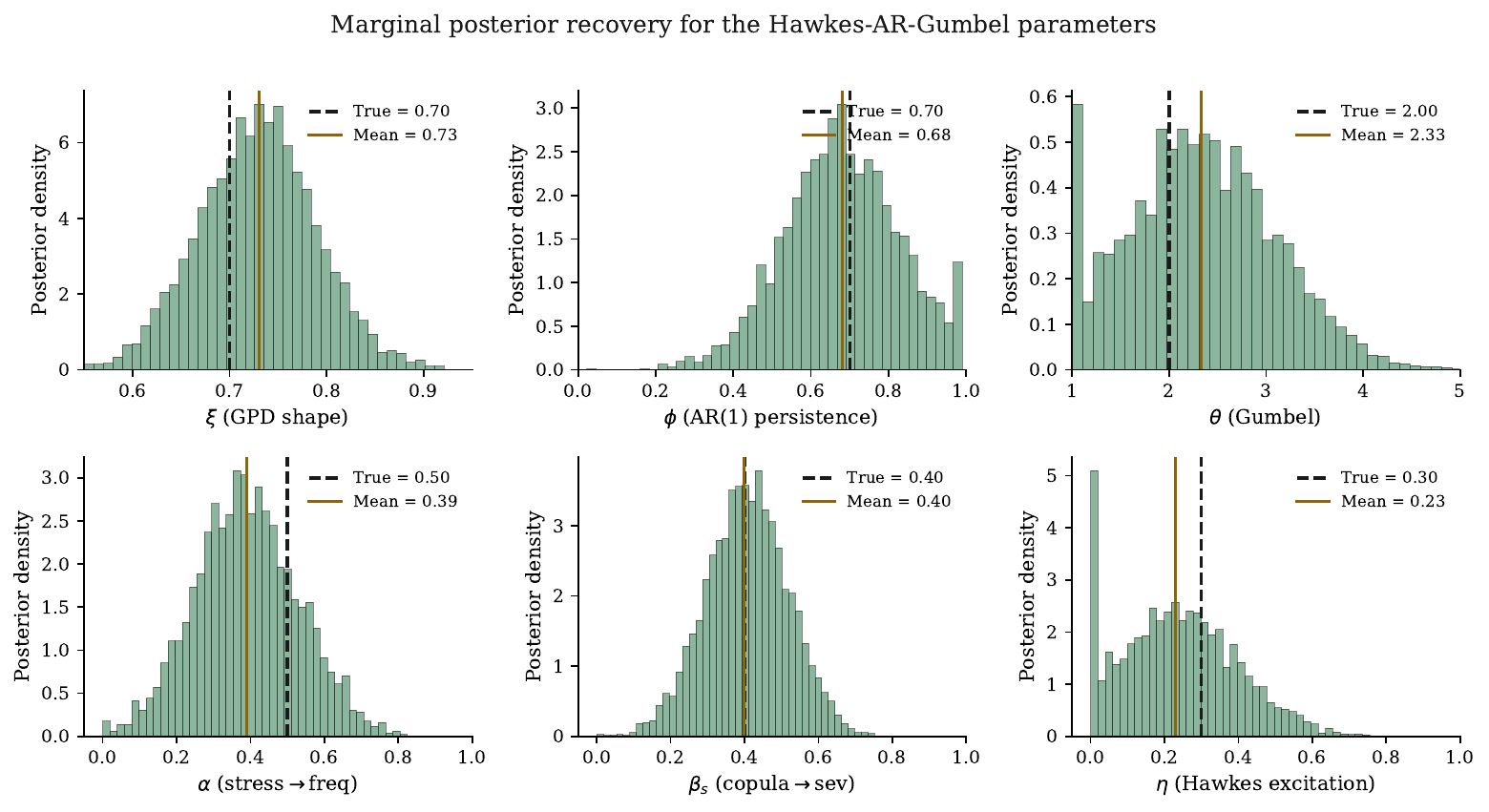}
\caption{Marginal posterior recovery of the structural parameters across the
  three fitted models, with the true data-generating values marked by dashed
  vertical lines. The Hawkes-AR-Gumbel posteriors concentrate around the
  true values; the simpler models cannot represent some of the parameters
  (notably $\theta$, $\eta$, $\kappa$, $\phi$) and therefore omit them.}
\label{fig:recovery}
\end{figure}

\begin{figure}[htbp]
\centering
\includegraphics[width=0.7\textwidth,keepaspectratio]{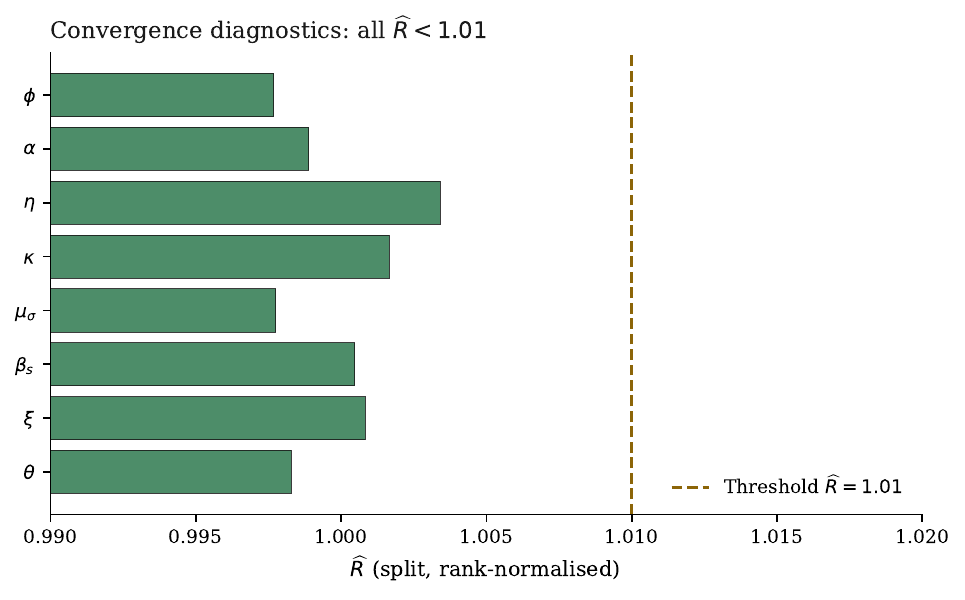}
\caption{Split-$\widehat{R}$ diagnostics for all monitored parameters across
  the three fitted models. All values are below the conventional threshold
  of~1.01, supporting convergence of the Hamiltonian Monte Carlo sampler.}
\label{fig:rhat}
\end{figure}

All true parameter values fall within the 94\% highest density intervals,
confirming that the model is well-calibrated. The severity parameters
$\mu_\sigma$, $\beta_s$, and $\xi$ are recovered with high precision, which
is expected given that approximately~925 exceedances provide substantial
information about the GPD. The persistence parameter $\phi$ is recovered
with a posterior mean of~0.68 against a true value of~0.70, and the copula
parameter $\theta$ has a posterior mean of~2.33 versus a true value
of~2.00, implying an estimated upper-tail dependence coefficient
$\widehat{\lambda}_U \approx 0.63$. The Hawkes parameters $\eta$ and
$\kappa$ show wider posteriors than the other parameters. This is to be
expected: with only $T = 15$ annual observations of event counts, it is
difficult to disentangle the self-excitation contribution from the
stress-driven variation in frequency. The posterior for $\eta$ places
non-negligible mass near zero, indicating that the data are consistent with
both substantial and negligible self-excitation. Nevertheless, both true
values lie within the posterior credible intervals, and the model does not
introduce systematic bias. Figure~\ref{fig:recovery} compares the marginal
posteriors against the data-generating values for all three fitted models,
and Figure~\ref{fig:rhat} reports the convergence diagnostics.

\subsection*{CVaR comparison across models}\label{sec:cvar_results}

Table~\ref{tab:cvar} presents the CVaR estimates from all three models,
computed via $10^6$ posterior predictive simulations per model.

\begin{table}[htbp]
\centering
\caption{CVaR estimates (in millions of monetary units) from posterior
  predictive simulation ($10^6$ draws per model). The ratio column shows
  the Hawkes-AR-Gumbel CVaR relative to the independent model.}
\label{tab:cvar}
\begin{tabular}{lrrrr}
\toprule
Level & Independent & Shared factor & Hawkes-AR-Gumbel & Ratio (HAG/Indep.) \\
\midrule
99.900\% &  37.9 &  35.6 &  43.9 & 1.16$\times$ \\
99.950\% &  63.6 &  59.8 &  74.3 & 1.17$\times$ \\
99.995\% & 322.3 & 326.3 & 461.4 & 1.43$\times$ \\
\bottomrule
\end{tabular}
\end{table}

The central result is that the independent model underestimates the CVaR
at~99.995\% by approximately~40\%, with a value of~322.3M compared
to~461.4M for the Hawkes-AR-Gumbel model. For a bank holding operational
risk capital calibrated to the independent model, this underestimation
would translate directly into an insufficient capital buffer, with the
shortfall growing more severe as the confidence level increases.

Perhaps more surprising is the performance of the shared latent factor
model. Despite capturing meaningful frequency-severity dependence through
the shared $Z_t$, its CVaR at~99.995\% (326.3M) is essentially
indistinguishable from the independent model's estimate. This failure is not
a numerical accident: it reflects the structural inability of the Gaussian
copula implicit in the shared factor model to generate upper-tail
dependence. At moderate quantiles (99.9\%), the three models produce similar
estimates, because at these levels the aggregate loss is driven primarily
by the mean behavior of frequency and severity, where all three models
agree. The divergence emerges only at the extreme quantiles, where the
aggregate loss is driven by the joint occurrence of extreme frequency and
extreme severity, a regime where the copula structure becomes decisive.
The growth pattern of the ratio across confidence levels is characteristic
of upper-tail dependence. At~99.9\%, the ratio is 1.16$\times$;
at~99.995\%, it reaches 1.43$\times$. This amplification occurs because the
Gumbel copula's effect operates through the joint extremes of the
innovation processes, which are rare events that contribute
disproportionately to the far tail of the aggregate loss distribution.

\subsection*{Mechanism decomposition}\label{sec:decomposition}

\begin{figure}[htbp]
\centering
\includegraphics[width=\textwidth,keepaspectratio]{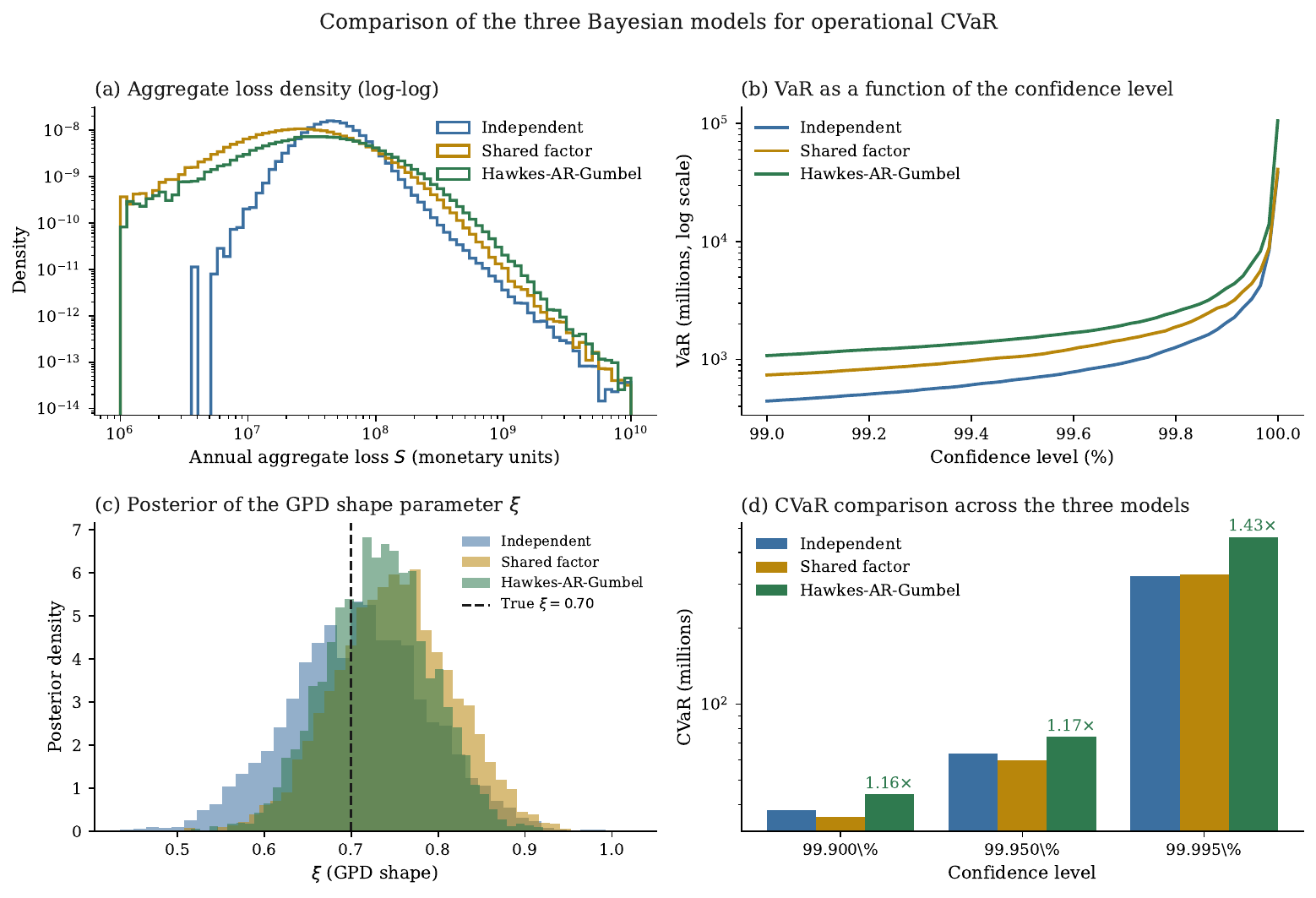}
\caption{Comparison of the three Bayesian models. (a) Aggregate loss
  distribution. (b)~VaR as a function of the confidence level (logarithmic
  axis). (c)~Posterior of the GPD shape parameter $\xi$.
  (d)~CVaR comparison at three reference confidence levels. The
  Hawkes-AR-Gumbel model (green) produces heavier tails and higher CVaR at
  extreme quantiles, while the shared factor model (orange) tracks the
  independent baseline (blue) at the most extreme quantile despite
  capturing intermediate dependence.}
\label{fig:results}
\end{figure}

The three components of the Hawkes-AR-Gumbel model contribute to the CVaR
increase through distinct channels that compound multiplicatively in the
tail. The upper-tail dependence introduced by the Gumbel copula ensures
that extreme frequency innovations are accompanied by extreme severity
innovations. With $\lambda_U \approx 0.6$, conditional on the frequency
innovation being in its 5\% upper tail, there is a $\sim$60\% probability
that the severity scale is also extreme. The Gaussian copula implicit in
the shared factor model has $\lambda_U = 0$, which means that simultaneous
extremes in frequency and severity become asymptotically independent,
producing a thinner aggregate tail. The temporal persistence introduced by
the AR(1) structure amplifies the unconditional variance of the stress
process. With $\phi = 0.7$, the unconditional variance of $Z_t$ is
$1/(1-\phi^2) \approx 1.96$, nearly twice the unit variance under
independent $Z_t \sim \Nor(0,1)$. This wider stress distribution increases
the probability of observing extended high-stress periods, which compound
the aggregate loss through multiple consecutive years of elevated frequency
and severity. The event clustering produced by the Hawkes self-excitation
amplifies the expected frequency by a factor of $1/(1-r)$, where the
posterior branching ratio $r$ ranges from approximately~0.3 to~0.5. This
creates overdispersion and temporal clustering beyond what Poisson variation
alone would generate, further inflating the right tail of the aggregate
loss distribution. Figure~\ref{fig:results} summarizes the comparison
across the three fitted models, displaying the aggregate loss distribution,
the VaR curves as a function of confidence level, the marginal posterior of
$\xi$, and the CVaR comparison at three reference quantiles.

\begin{figure}[htbp]
\centering
\includegraphics[width=\textwidth,keepaspectratio]{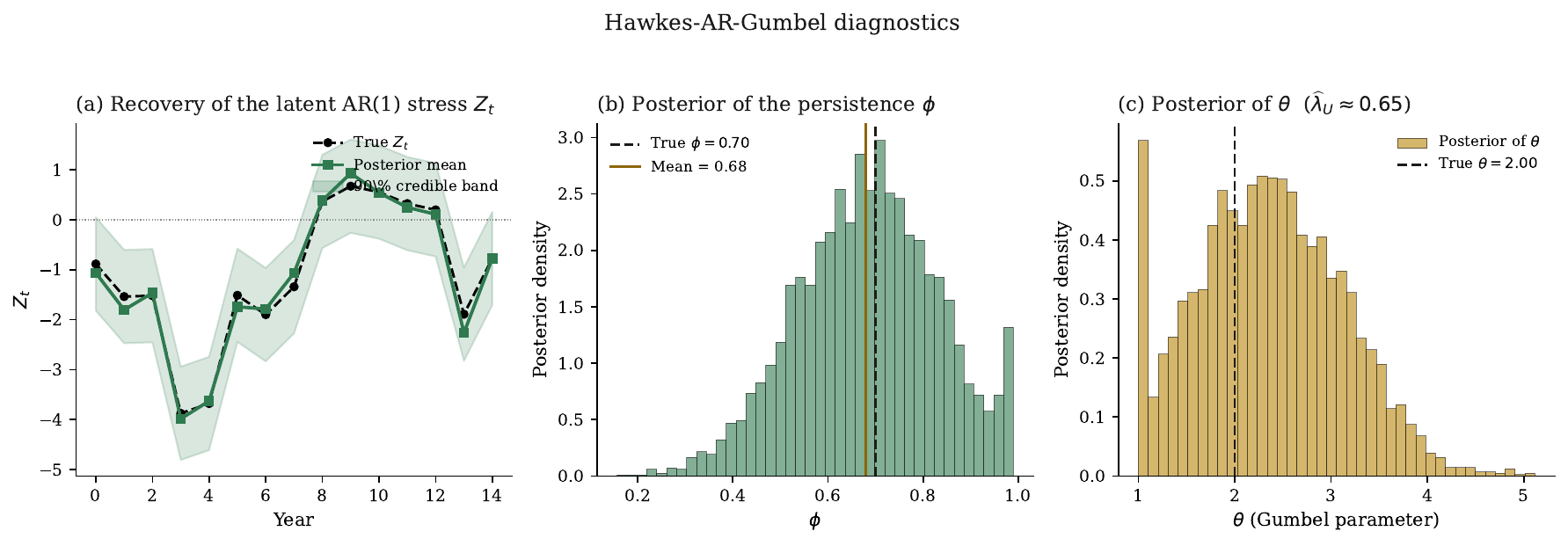}
\caption{Diagnostics for the Hawkes-AR-Gumbel model. (a)~Recovery of the
  latent AR(1) stress process $Z_t$ (posterior mean and 90\% credible band
  versus the true values). (b)~Posterior of the persistence parameter
  $\phi$. (c)~Posterior of the Gumbel copula parameter $\theta$, with the
  implied upper-tail dependence coefficient $\lambda_U$.}
\label{fig:diagnostics}
\end{figure}

\begin{figure}[htbp]
\centering
\includegraphics[width=\textwidth,keepaspectratio]{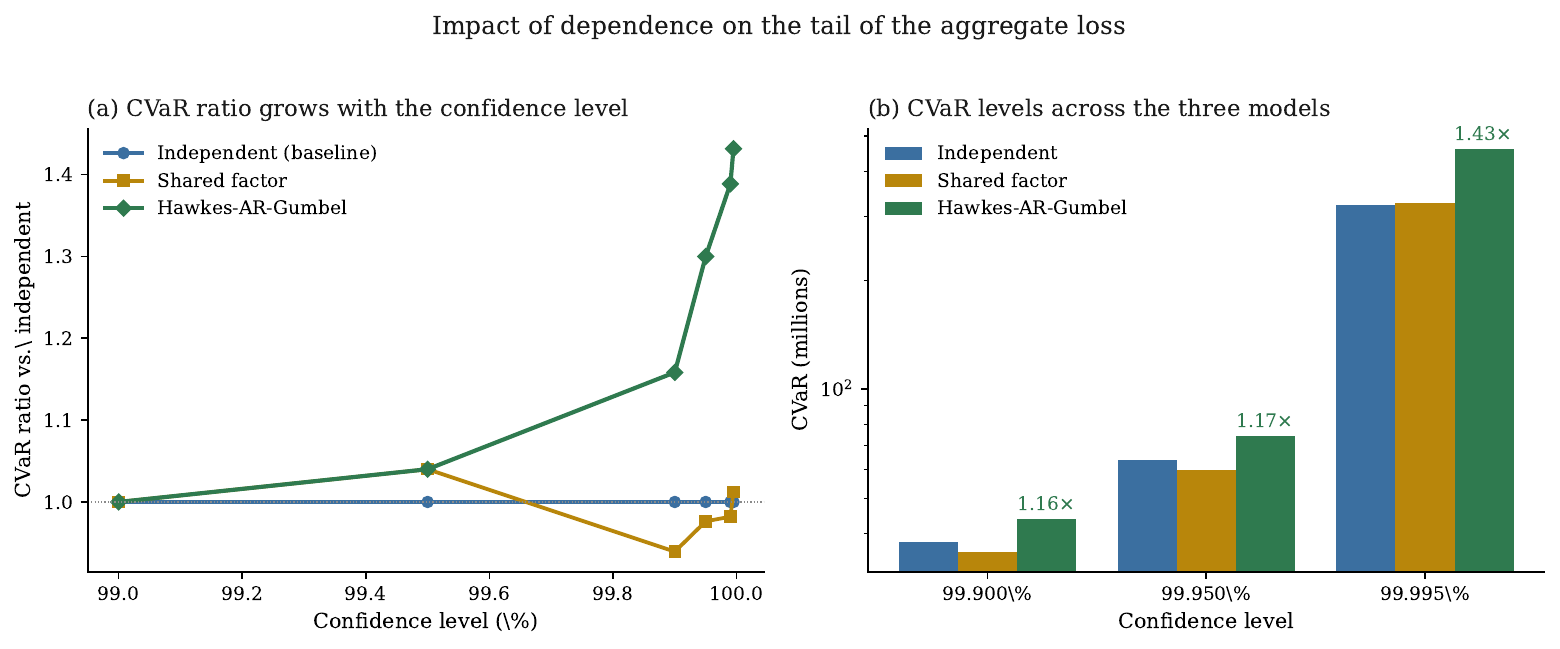}
\caption{Impact of dependence on the tail of the aggregate loss
  distribution. Each panel shows the increase in VaR and CVaR as the
  dependence structure is enriched, illustrating that the gap relative to
  the independent baseline grows with the confidence level. The plot
  visualizes the multiplicative compounding of the three Hawkes-AR-Gumbel
  mechanisms in the extreme tail.}
\label{fig:dependence}
\end{figure}

\section{Discussion}\label{sec:discussion}

The finding that the standard LDA underestimates CVaR at~99.995\% by
approximately~40\% has direct implications for regulatory capital adequacy.
Under Basel~II, banks using the Advanced Measurement Approach (AMA) were
permitted to use internal models for operational risk capital, subject to
supervisory validation. While Basel~III has moved toward the Standardized
Measurement Approach (SMA), the internal model insights remain relevant
for Pillar~2 stress testing, where banks must demonstrate adequate capital
under adverse scenarios; for the Internal Capital Adequacy Assessment
Process (ICAAP), where internal models inform capital planning; and for
economic capital allocation across business lines. In all three contexts, a
model that systematically underestimates tail risk provides a misleading
basis for decision-making.

The failure of the shared Normal factor model to improve CVaR at extreme
quantiles, despite capturing meaningful frequency-severity dependence at
moderate levels, is a cautionary result that deserves emphasis. The shared
factor model is not a naive baseline: it correctly identifies that stress
increases both frequency and severity, and its posterior estimates of the
sensitivity parameters $\alpha$ and $\beta$ are statistically significant.
Yet its CVaR at~99.995\% is essentially identical to the independent
model. The explanation lies in the Gaussian copula's zero upper-tail
dependence coefficient. As the confidence level increases, the CVaR becomes
increasingly sensitive to the probability of simultaneous extremes in
frequency and severity. Under the Gaussian copula, this probability
vanishes faster than under the Gumbel copula, so the apparent dependence
captured at moderate quantiles fails to translate into improved tail
estimation. This result illustrates a broader lesson for risk modeling: the
type of dependence matters as much as its strength, and fitting a model
that matches central moments or moderate quantiles does not guarantee
accurate tail behavior.

The practical importance of frequency-severity dependence depends on two
factors. The first is the confidence level. At moderate quantiles below
the~99\% level, all three models produce similar CVaR estimates, because at
these levels the aggregate loss is driven by average behavior where all
models approximately agree. The divergence emerges above~99.9\% and grows
with the confidence level, because the copula effect operates through the
joint occurrence of extreme innovations, which are rare events that only
influence the far tail. The second factor is the tail heaviness
parameter~$\xi$. For lighter-tailed distributions ($\xi < 0.5$), the impact
of dependence on CVaR is attenuated because individual extreme severities
are bounded. For the heavy tails typical of operational risk
($\xi > 0.5$), the combination of extreme frequency and extreme severity
can produce catastrophic aggregate losses, making the dependence structure
economically consequential.

Our study has several limitations that point toward future research. First,
our results are based on simulated data from a known DGP, which guarantees
parameter recovery but does not demonstrate real-world applicability.
Application to internal loss databases, subject to the confidentiality
constraints that limit publication of bank-specific operational risk data,
is the natural next step. Second, we model a single operational risk cell
(one business line by one event type); extension to multiple cells with
cross-cell dependence is conceptually straightforward but computationally
demanding. Third, the Hawkes component operates at annual granularity,
while the theoretical Hawkes process is continuous-time; sub-annual data,
where available, would allow a richer self-excitation structure. Fourth, we
compare models visually and through CVaR ratios; formal Bayesian model
comparison via WAIC or LOO-CV \citep{vehtari2017} would strengthen the
analysis. Finally, our normal marginal innovations could be replaced with
skew-$t$ distributions \citep{garzon2023} to capture asymmetric stress
shocks, an extension that is straightforward within the copula framework
but adds two additional parameters per innovation.

\section{Conclusion}\label{sec:conclusion}

We have proposed the Hawkes-AR-Gumbel model, a Bayesian framework for
operational risk CVaR estimation that integrates three mechanisms absent
from the standard Loss Distribution Approach: upper-tail dependence between
frequency and severity via the Gumbel copula, temporal persistence of
stress regimes via an AR(1) latent process, and event clustering via
Hawkes self-excitation. On simulated data, the model correctly recovers the
data-generating process parameters and demonstrates that ignoring these
mechanisms leads to substantial underestimation of CVaR at extreme
quantiles, approximately~40\% at the~99.995\% level. The shared Normal
latent factor model, despite capturing symmetric frequency-severity
dependence, fails to improve CVaR estimation at extreme quantiles because
the Gaussian copula lacks upper-tail dependence. This finding underscores a
key message for practitioners: at the extreme quantiles relevant for
regulatory capital ($\geq 99.9\%$), the type of dependence matters as much
as its presence. Symmetric dependence structures capture average co-movement
but miss the asymmetric synchronization of extremes that drives
catastrophic aggregate losses. The Gumbel copula provides this missing
piece with minimal additional computational cost. The framework is
computationally tractable, with MCMC inference completing in under~15
minutes on a standard laptop; fully Bayesian, providing honest uncertainty
quantification at quantiles where frequentist approximations break down;
and modular, with each component (copula, AR persistence, Hawkes
excitation) activatable independently for model comparison. We provide a
complete open-source implementation in Python and PyMC to facilitate
reproducibility and adoption by risk management practitioners.

\appendix

\section{Gumbel Copula Sampling Algorithm}\label{app:gumbel}

Sampling from the bivariate Gumbel copula $C_\theta$ is performed using the
Marshall-Olkin algorithm \citep{nelsen2006}. The algorithm exploits the
representation of Archimedean copulas through a latent frailty variable:
given a positive stable random variable $M$ with Laplace transform
$\E[e^{-tM}] = e^{-t^a}$, where $a = 1/\theta$, the pair
$(U, V) = (\exp(-(E_1/M)^a), \exp(-(E_2/M)^a))$ with
$E_1, E_2 \iid \text{Exp}(1)$ follows the Gumbel copula $C_\theta$. The
positive stable variate $M$ is generated using the Chambers-Mallows-Stuck
algorithm. One draws $V \sim \text{Uniform}(-\pi/2, \pi/2)$ and
$W \sim \text{Exp}(1)$, then computes
\begin{equation}
  M = \frac{\sin(a(V + \pi/2))}{(\cos V)^{1/a}} \cdot
  \left(\frac{\cos(V - a(V + \pi/2))}{W}\right)^{(1-a)/a}.
\end{equation}
This algorithm is exact and numerically stable for $a \in (0, 1)$, which
corresponds to $\theta > 1$. The boundary case $\theta = 1$ recovers the
independence copula and is handled separately by drawing $U$ and $V$
independently uniform.

\section{Gumbel Copula Log-Density}\label{app:density}

The log-density of the bivariate Gumbel copula, used in the MCMC
likelihood (equation~\eqref{eq:loglik}), takes the form
\begin{align}
  \log c_\theta(u, v) &= -A^{1/\theta}
    + \log u + \log v
    + (\theta - 1)(\log\ell_u + \log\ell_v)
    \nonumber\\
    &\quad - \left(2 - \frac{1}{\theta}\right) \log A
    + \log\!\left(A^{1/\theta} + \theta - 1\right),
\end{align}
where $\ell_u = -\ln u$, $\ell_v = -\ln v$, and
$A = \ell_u^\theta + \ell_v^\theta$. In the PyMC implementation, the
copula correction is added as a \texttt{pm.Potential} to the log-posterior.
The innovation variables $W_t^f$ and $W_t^s$ are given independent
$\Nor(0,1)$ priors, and the potential adds the copula log-density evaluated
at $u = \Phi(W_t^f)$, $v = \Phi(W_t^s)$. The combined effect is that the
joint posterior for $(W_t^f, W_t^s)$ has the correct Gumbel copula
structure with standard normal marginals, without requiring direct sampling
from the copula within the MCMC.

\bibliographystyle{plainnat}
\bibliography{references}

\end{document}